\documentclass[%
 aip,
 amsmath,amssymb,
 reprint,%
]{revtex4-1}

\usepackage{graphicx}% Include figure files
\usepackage{dcolumn}% Align table columns on decimal point
\usepackage{bm}% bold math
\usepackage[utf8]{inputenc}
\usepackage[T1]{fontenc}
\usepackage{mathptmx}
\usepackage{xcolor} %required for \textcolor command

\begin{document}

\preprint{AIP/123-QED}

\title{Room temperature cavity electromechanics in the sideband-resolved regime}
% Force line breaks with \\

\author{Anh Tuan Le}
\affiliation{ 
	Department of Physics, University of Konstanz, 78457 Konstanz, Germany%\\This line break forced with \textbackslash\textbackslash
}%
\affiliation{ 
	Department of Electrical \& Computer Engineering, Technical University of Munich, 80333 Munich, Germany%\\This line break forced with \textbackslash\textbackslash
}%
\author{A. Brieussel}%
\affiliation{ 
	Department of Physics, University of Konstanz, 78457 Konstanz, Germany%\\This line break forced with \textbackslash\textbackslash
}%
\author{E.~M. Weig}
\email{eva.weig@tum.de}
\affiliation{ 
	Department of Physics, University of Konstanz, 78457 Konstanz, Germany%\\This line break forced with \textbackslash\textbackslash
}%
\affiliation{ 
	Department of Electrical \& Computer Engineering, Technical University of Munich, 80333 Munich, Germany%\\This line break forced with \textbackslash\textbackslash
}%
\affiliation{ 
	Munich Center for Quantum Science and Technology (MCQST), 80799 Munich, Germany%\\This line break forced with \textbackslash\textbackslash
}%

\date{\today}% It is always \today, today,

\begin{abstract}
 We demonstrate a sideband-resolved cavity electromechanical system operating at room temperature. It consists of a nanomechanical resonator, a strongly pre-stressed silicon nitride string, dielectrically coupled to a three-dimensional microwave cavity made of copper. The electromechanical coupling is characterized by two measurements, the cavity-induced eigenfrequency shift of the mechanical resonator and the optomechanically induced transparency. While the former is dominated by dielectric effects, the latter reveals a clear signature of the dynamical backaction of the cavity field on the resonator. This unlocks the field of cavity electromechanics for room temperature applications. 
\end{abstract}

\maketitle
\section{Introduction}
In recent years, cavity (or circuit) electromechanics has been established as a powerful implementation of cavity optomechanics.~\cite{Marquardt2014} Rather than relying on a light field circulating inside an optical cavity, an electromagnetic circuit is employed to realize a cavity mode in the microwave regime which parametrically couples to a mechanical resonator.~\cite{Regal2008,Teufel2008,Teufel2011,Pernpeintner2014a}    
This seemingly simple modification has enabled a series of breakthroughs, including
quantum ground state cooling,~\cite{Teufel2011a} 
entanglement generation,~\cite{Palomaki2013a,Barzanjeh2019}
%non-classical state generation, i.e. squeezing: first observed with cOM (by a few weeks)~\cite{Purdy2013,Safavi-Naeini2013}
long-lived quantum storage,~\cite{Palomaki2013,Zhou2013,Pechal2018}
microwave-to-optical conversion,~\cite{Andrews2014,Forsch2019} non-reciprocal signal transduction,~\cite{Barzanjeh2017} 
%strong coupling first observed with cOM~\cite{Groeblacher2009}
and ultrastrong coupling~\cite{Peterson2019}, just to name a few. Following recent developments in the field of superconducing qubits~\cite{Paik2011,Reagor2013} where three-dimensional microwave cavities have replaced coplanar waveguide architectures for their large mode volumes and remarkably high quality factors, three-dimensional superconducting microwave cavities have been adapted for cavity electromechanics by capacitive coupling to a mechanical resonator.~\cite{Yuan2015,Noguchi2016,Carvalho2019}
However, to date, the field of cavity electromechanics is limited to millikelvin temperatures, since it relies on superconducting circuits. Room temperature cavity electromechanics is impeded by the non-zero resistance of normal conducting circuits which gives rise to dissipation. Non-superconducting microwave cavities such as copper microstrip resonators have been successfully employed for cavity-assisted displacement sensing of nanomechanical resonators at room temperature.~\cite{Faust2012,Rieger2012} Their use for cavity electromechanics~\cite{Faust2012} is limited by a cavity quality factor of about $100$ due to dielectric and conductor losses, which not only constrains the displacement sensitivity but also keeps the system deeply in the unresolved sideband, so-called bad cavity regime where the linewidth of the cavity $\kappa/2\pi$ exceeds the frequency of the mechanical mode $\Omega_m/2\pi$.

Here we present a room temperature cavity electromechanical system capable of sideband-resolved operation ($\kappa < \Omega_m$). Inspired by the developments in the field of superconducing qubits~\cite{Paik2011,Reagor2013} and cavity electromechanics~\cite{Yuan2015,Noguchi2016,Carvalho2019} we employ a three-dimensional, cylindrically shaped cavity rather than a microstrip to dielectrically probe the displacement of a strongly pre-stressed silicon nitride nanostring resonator. 

\section{Results and discussion}
\begin{figure*}[hbt!]
	\includegraphics{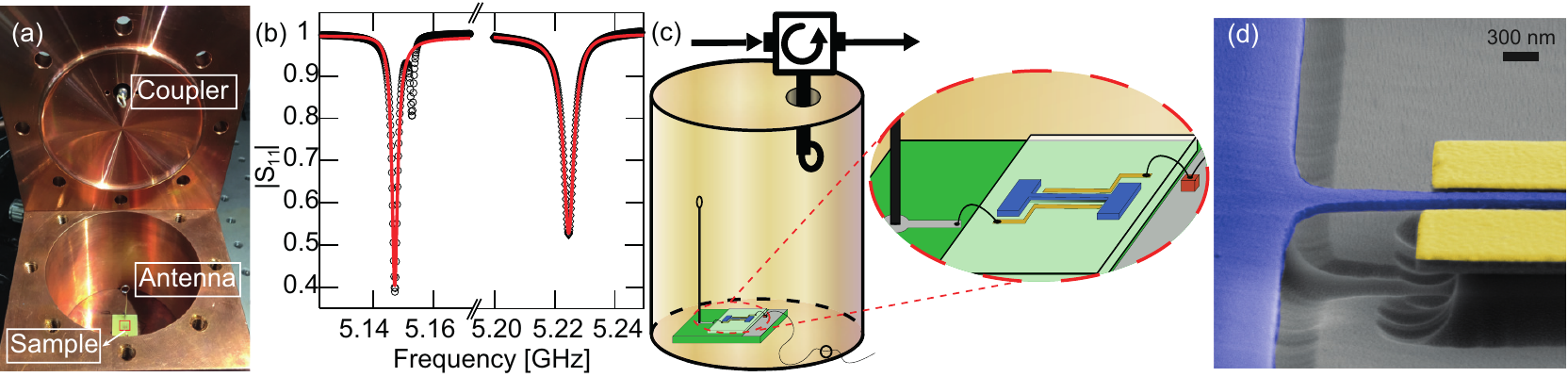}% Here is how to import EPS art
	\caption{\label{fig:setup}\textbf{Experimental setup of the cavity electromechanical system at room temperature.} \textbf{(a)}, Photograph of the (open) cylindrical microwave 3D cavity machined from bulk copper. The microwave signal is inductively injected into the cavity via a loop coupler. The sample holder including the antenna is visible on the cavity floor. \textbf{(b)}, Reflection coefficient $\left|S_{11}\right|$ of the $\textrm{TM}_{110}$ mode (black) along with fit to the data (red), showing the response of both the empty cavity (left) and the cavity including the sample (right). Upon insertion of the sample the frequency of the $\textrm{TM}_{110}$ mode shifts by $\sim 80$\,MHz. Furthermore, an increase of the linewidth by a factor of $2$ is observed. \textbf{(c)}, Schematic visualizing the physical implementation of the cavity electromechanical system. The doubly clamped silicon nitride string resonator (blue) is situated between two elevated gold electrodes (yellow). The resonator chip (light green) is glued to a printed circuit board (green) and placed inside the cavity (beige). The circuit board also hosts a looped dipole antenna which is bonded to one of the electrodes to inductively couple to the $\textrm{TM}_{110}$ mode. The other electrode is connected to a single layer capacitor (brown) which serves as a capacitive ground for frequencies in the microwave frequency range, and to a wire which is fed through a hole in the cavity wall. This allows to apply DC voltages and RF signals to dielectrically drive the resonator inside the cavity. \textbf{(d)}, Scanning electron micrograph of a doubly clamped pre-stressed nanomechanical resonator between two gold electrodes for dielectric actuation as well as coupling to the three-dimensional microwave cavity. 
	}
\end{figure*}

Figure~\ref{fig:setup}(a) depicts a photograph of the cylindrical cavity. It consists of two parts which have been machined from bulk copper (Cu > 99.90 \%) and which can be closed using screws. The cavity has a radius of $35$ mm and a height of $70$ mm, it supports both transverse electric (TE) and transverse magnetic (TM) modes.~\cite{Pozar2011} As the modes reside in the hollow cylinder, they are only weakly affected by dielectric and conductor losses, enabling high quality factor even at room temperature.~\cite{Reagor2013} The coupling to the cavity is realized by injecting the microwave signal from a coaxial line through a loop coupler via a $3.6$ \,mm sized hole in the cavity top. In all our measurements, we use a circulator to physically separate the input and output signals of the single port reflection cavity.
% and to avoid back reflections and interference effects. 
Figure \ref{fig:setup}(b) (left) displays the reflection coefficient $\left|S_{11}\right|$ of the $\textrm{TM}_{110}$ mode which is found at $\omega_c/(2\pi) \approx 5.147$\,GHz with a linewidth of $\kappa/(2\pi) = 2.824$\,MHz for the empty cavity, in good agreement with finite element simulations. This corresponds to a cavity quality factor of $1,800$, exceeding the state of the art in microwave-cavity-assisted nanomechanical displacement sensing at room temperature by more than an order of magnitude.~\cite{Faust2012} 
The nanomechanical resonator under investigation is a strongly pre-stressed nanostring fabricated from LPCVD silicon nitride on a fused silica wafer, which is flanked by two gold electrodes for dielectric transduction.~\cite{Faust2012,Rieger2012} The reflection coefficient of the cavity including the resonator chip is shown in Figure \ref{fig:setup}(b) (right). Upon insertion of the resonator, the eigenfrequency of the  $\textrm{TM}_{110}$ mode shifts to $\omega_c/(2\pi) = 5.226$\,GHz. The linewidth increases by a factor of $2$ to $\kappa/(2\pi) = 5.572$\,MHz.

\begin{figure}[hbt!]
	\includegraphics{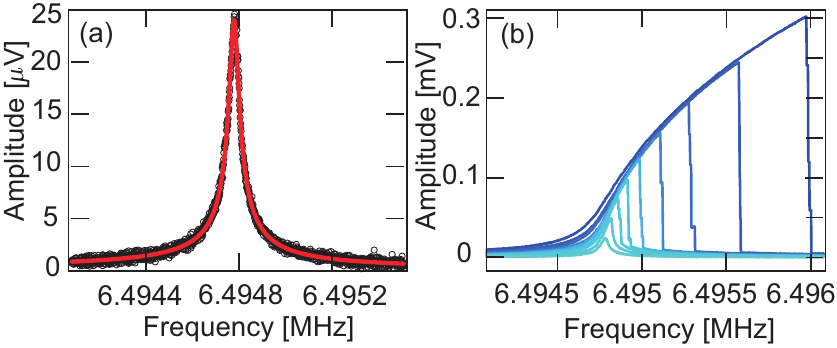}% Here is how to import EPS art
	\caption{\label{fig:Fig_2} \textbf{Characterization of the nanostring resonator.} \textbf{(a)}, Linear response of the fundamental out-of-plane mode of the nanostring (black) and Lorentzian fit (red) for a drive power of $P_{m}=-36$\,dBm. \textbf{(b)}, Response of the nanostring by increasing drive power from $P_{m}=-36$\,dBm to $-14$\,dBm showing the transition to the nonlinear Duffing regime.
	}
\end{figure}

%By inserting the sample into the cavity the resonance frequency shifts from $\omega_c/(2\pi) = 5.147 $ GHz to $\omega_c/(2\pi) = 5.2267 $ GHz and the cavity linewidth is broadened by $\kappa/(2\pi) = 2.824 $ MHz to $\kappa/(2\pi) = 5.005 $ MHz
In addition, the fit of the amplitude and phase of the cavity's reflection coefficient reveals an external dissipation rate of $\kappa_{ex}/(2\pi)= 1.361$\,MHz. This gives rise to a coupling efficiency $\eta = \kappa_{ex}/\kappa=0.244\ $, indicating that the cavity is undercoupled.  Figure~\ref{fig:setup}(c) shows the physical realization of the cavity electromechanical system. The sample holder with the resonator chip is placed inside the cavity. One of the electrodes is connected to an antenna to inductively couple to the $\textrm{TM}_{110}$ of the cavity. It consists of a looped silver wire on top of a coaxial cable which places the loop into the electromagnetic field near the center of the hollow cylinder. The other electrode is bonded across a single layer capacitor acting as capacitive ground~\cite{Rieger2012} and connected to an RF signal generator through a $1.5$\,mm wide hole in the cavity wall for dielectric actuation of the resonator.~\cite{Chen2011,Graaf2014,Hao2014a,Kong2015,Cohen2017,Stammeier2018}
The nanostring under investigation is $w=250$\,nm wide, $t=100$\,nm thick and $L=57\,\mu$m long, similar to the one depicted in Fig.~\ref{fig:setup}(d). 
%
%******************
%The eigenfrequency of its fundamental out-of-plane mode is
%
%\begin{eqnarray}
%\Omega_m=2\pi\times\frac{\pi}{2L^2}\sqrt{\frac{EI}{\rho A}}\sqrt{1+\frac{\rho AL^2}{EI\pi^2}}
%\end{eqnarray}
%
%according to Euler-Bernoulli beam theory~\cite{Weaver1990}, where $ L $ denotes the length , $ E $ the Young’s modulus, $ \rho $ the %density and $ \sigma $ the tensile pre-stress of the nanomechanical resonator. The area moment of inertia of the out-of-plane mode is %$I = wt^3/12$. 
%******************
%
While the cavity characterization is done under ambient conditions, the experiments discussed in the following involve the vibrational excitation of the nanostring and air damping needs to be excluded. Hence, the entire cavity electromechanical system consisting of the string resonator as well as the microwave cavity is placed inside a vacuum chamber below $5\times 10^{-4}$\,mbar. As indicated above, all measurements are performed at room temperature.
The nanomechanical resonator is characterized by dielectrically driving its fundamental out-of-plane mode with a vector network analyzer (VNA). The response of the nanomechanical resonator is characterized using heterodyne cavity-assisted displacement detection~\cite{Faust2012} using the three-dimensional microwave cavity. The microwave cavity is driven on resonance $\omega_d = \omega_c$, such that the periodic modulation of the capacitance induced by the dielectric nanostring vibrating between the two electrodes induces sidebands at $\omega_c\pm\Omega_m $ on the cavity response. The reflected signal of the cavity is demodulated by an in-phase quadrature (IQ) mixer, filtered and amplified as previously described.~\cite{Faust2012} The resulting signal is fed back into the VNA. The response of the fundamental out-of-plane mode of the resonator is displayed in Fig.~\ref{fig:Fig_2}. Figure~\ref{fig:Fig_2}(a) shows the Lorentzian resonance curve observed in the linear response regime for a drive power $P_m=-36$\,dBm. Fitting allows to extract the resonance frequency $\Omega_m/(2\pi)= 6.4948$\,MHz, as well as a linewidth $\Gamma/(2\pi) = 42$\,Hz which gives rise to a quality factor of $Q\approx 150.000$. Given the cavity linewidth $\kappa/(2\pi) = 5.572$\,MHz the condition for sideband resolution, $\Omega_m > \kappa$, is fulfilled for the $\textrm{TM}_{110}$ mode. The response of the resonator for increasing drive power between $-36$\,dBm to $-14$\,dBm is depicted in Fig.~\ref{fig:Fig_2}(b). Clearly, the transition from the linear, Lorentzian response to an asymmetric response curve, which is well described by the cubic nonlinearity $\alpha x^3$ of the Duffing model with a stiffening $\alpha>0$, is observed.~\cite{Nayfeh1995} This demonstrates that the cavity-assisted displacement detection is not impeded even under strong driving of the nanomechanical resonator inside the 3D cavity.

Following the characterization of the microwave cavity and the resonator, we discuss the electromechanical coupling between the two systems. First, we explore how the cavity detuning affects the mechanical eigenfrequency of the resonator. The radiation pressure of the power circulating in the microwave cavity acts back on the mechanical resonator, causing a mechanical eigenfrequency shift which depends on the microwave drive power $P_d $ and detuning $\Delta = \omega_d - \omega_c  $. This so-called optical spring effect leads to a detuning-dependent softening or hardening of the resonator. In case of a high-Q mechanical oscillator with small linewidth, where $ \Gamma\ll\kappa$, the frequency shift of the mechanical oscillator can be defined as~\cite{Marquardt2014}
\begin{eqnarray}\label{eq:1}
\delta\Omega_{m,\textrm{opt}} = g^2\left(\frac{\Delta-\Omega_m}{\frac{\kappa^2}{4}+\left(\Delta-\Omega_m\right)^2}+\frac{\Delta+\Omega_m}{\frac{\kappa^2}{4}+\left(\Delta+\Omega_m\right)^2} \right) .
\end{eqnarray}
The electromechanical coupling strength $g = g_0\cdot\sqrt{n_d}$ depends on the single photon coupling rate $g_0$ and the number of photons circulating in the cavity $n_d$. Note that this quantity strongly depends on the detuning of the cavity. Hence, for a given power $P_d$ the intra-cavity photon number 
$n_d=P_d \kappa_{ex}/\left[\hbar\omega_d \left( \frac{\kappa^2}{4}+\Delta^2\right)\right]$
%$n_d = \frac{\kappa_{ex}}{\frac{\kappa^2}{4}+\Delta^2}\frac{P_d}{\hbar\omega_d}$ 
of the single-sided cavity can be significantly reduced for a finite cavity detuning. 
As the static mechanical displacement of the resonator arising from the radiation pressure force~\cite{Marquardt2014} is negligibly small, the detuning $\Delta$ rather than the effective detuning $\bar\Delta$ is employed in eq.~(\ref{eq:1}) \\
%The quantity $\bar{\Delta}=\Delta +G\bar{x}$ is an effective cavity detuning including the frequency shift due the static mechanical displacement of the resonator~\cite{Shevchuk2015}.

\begin{figure}[hbt!]
\includegraphics{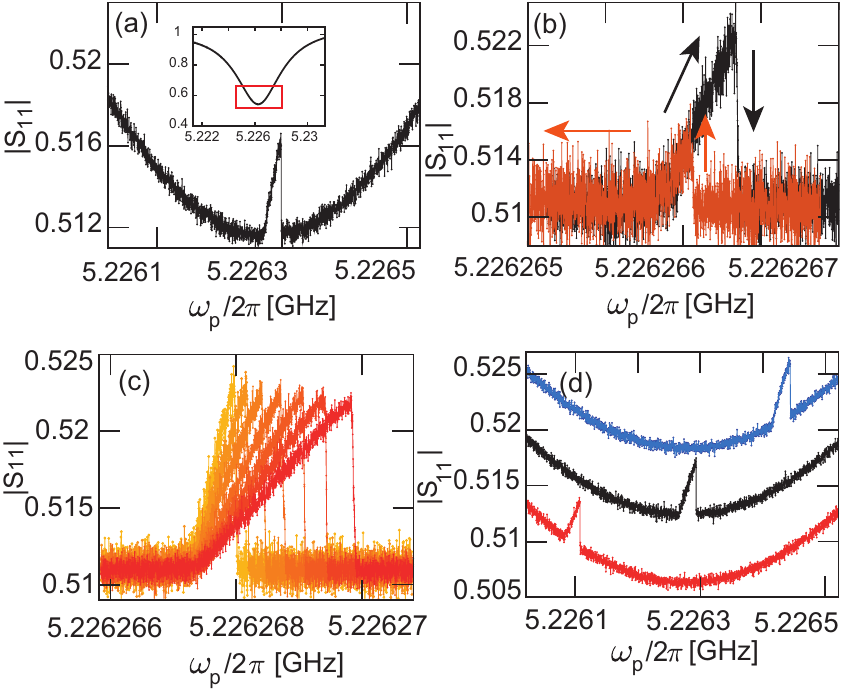}% Here is how to import EPS art
\caption{\label{fig:Fig_4} \textbf{Optomechanically induced reflection. (a)}, Cavity reflection coefficient $\left|S_{11}\right|$ showing an OMIR peak of the undercoupled cavity in response to a weak probe tone $\omega_p\approx\omega_c$ in the presence of an additional strong, red-detuned drive $\omega_d = \omega_c-\Omega_{m}$. The asymmetric shape of the OMIR feature reveals the nonlinearity of the mechanical resonator. The inset displays the full cavity resonance (red box illustrates area shown in main panel). \textbf{(b)}, Forward (blue) and reverse (red) sweeps of the probe tone reveal a hysteresis of the OMIR feature reflecting the bistability of the nonlinear mechanical system. \textbf{(c)} OMIR feature as a function of the probe power. For increasing probe power, the OMIR feature broadens, while its amplitude slightly decreases. \textbf{(d)} OMIR feature as a function of the drive detuning. For a drive tone red- or blue-shifted from the red sideband condition (red and blue trace, respectively) the OMIR feature shifts to the left or right of the cavity resonance. The situation for a drive on the red sideband is also included (black trace). Red and blue trace are vertically offset for clarity.
}
\end{figure}

In addition to the optomechanical backacktion we also expect a quasi-static dielectric force acting on the mechanical resonator. This force results from the root-mean-square (RMS) electrical field which builds up inside the microwave cavity and contributes to the dielectric frequency tuning of the nanostring~\cite{Rieger2012}. As the RMS cavity field exhibits the same detuning-dependence as the intracavity photon number $n_d$, this translates into a change in eigenfrequency of the mechanical oscillator as the cavity is detuned from its resonance
\begin{eqnarray}\label{eq:2}
\delta\Omega_{m,\textrm{dielectric}} = c_{\mu w}\cdot\left(U_{\mu w}^{rms}\left(\Delta\right)\right)^2 = c_{\mu w}R\frac{\kappa\kappa_{ex}}{\frac{\kappa^2}{4}+\Delta^2}P_d .
\end{eqnarray}  
The calibration factor $c_{\mu w}$ converts the change of the effective RMS voltage $U_{\mu w}^{rms}$ inside the cavity into a frequency change, and $R$ is the impedance of the circuit allowing to express the RMS voltage in terms of the power circulating in the cavity. In total, the detuning dependence of the mechanical resonance is
\begin{eqnarray}\label{eq:3}
\Omega'_m = \Omega_m+\delta\Omega_{m,\textrm{dielectric}}+\delta\Omega_{m,\textrm{opt}}
\end{eqnarray}
where $\Omega_m$ is the unperturbed mechanical eigenfrequency. 
\begin{figure}[t!]
	\includegraphics{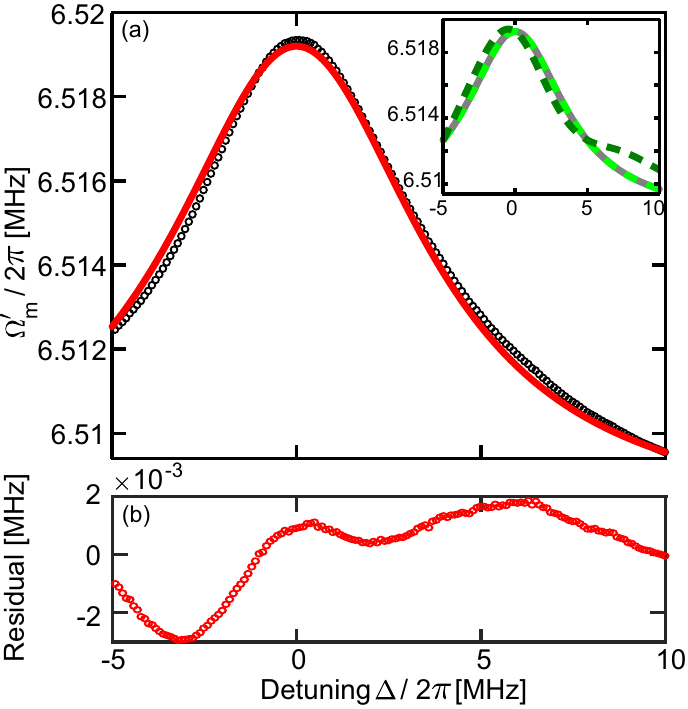}% Here is how to import EPS art
	\caption{\label{fig:Fig_3} \textbf{Cavity-induced eigenfrequency shift of the resonator for a cavity drive power of $P_d = 15$\,dBm. (a)}, Experimental data is shown as black circles, the fit of Eq.~(\ref{eq:3}) is indicated as a red solid line. The observed detuning dependence of the mechanical eigenfrequency is dominated by the dielectric shift, which overwhelms the contribution of the optomechanical backaction. Inset: Theoretical prediction of the eigenfrequency shift according to eq. (3) for $g_0=0$\,Hz (grey), $200\,\mu$Hz (light green) and $20$\,mHz (dark green). \textbf{(b)} Residual of the data shown in (a).}
\end{figure}
In Fig.~\ref{fig:Fig_3}(a) the nanomechanical resonator's eigenfrequency is plotted as a function of the detuning (black circles) for a cavity drive power of $P_d = 15$\,dBm. The data is well described by eq.~(\ref{eq:3}), which is apparent from a fit to the data (red line). However, the fit yields an overwhelming contribution of the dielectric frequency shift and only negligible optomechanical backaction such that the optomechanical coupling strength $g$ can not be determined. 
The inset of Fig.~\ref{fig:Fig_3}(a) shows the theoretical prediction of the eigenfrequency according to eq.~(\ref{eq:3}) using the parameters of the experiment for three different values of $g_0$: The grey solid line illustrates the eigenfrequency shift for zero optomechanical coupling $g_0 = 0$\,Hz. As eq.~(\ref{eq:2}) does not depend on the sign of $\Delta$, it is mirror symmetric with respect to the $y$-axis. The light green dashed line corresponds to $g_0=200\,\mu$Hz, which is the value of the single photon coupling rate estimated from the data in Fig.~\ref{fig:Fig_4}. The line completely coincides with the grey line, confirming that the optomechanical coupling has no measurable effect on the eigenfrequency shift. Minute deviations start becoming apparent from $g_0=2$\,mHz which we thus estimate as an upper bound of the single photon coupling rate. The dark green dashed line for $g_0=20$\,mHz already shows a sizable deviation of the eigenfrequency shift resulting from optomechanical coupling.
Our observation of a negligible optomechanical eigenfrequency shift is supported by the nearly mirror-symmetric shape of the measured eigenfrequency shift of the resonator around zero detuning in Fig.~\ref{fig:Fig_3}(a). The slight deviation shifting the maximum to a small positive detuning is likely caused by a slight drift of the mechanical eigenfrequency from slow polarization effects within the nanostring during the cavity frequency sweep, the opposite effect would be expected from radiation pressure effects. 
Figure Fig.~\ref{fig:Fig_3}(b) shows the residual of the data vs. the fit. We attribute the observed pattern to the same slow polarization effects.
%Based on eq.~(\ref{eq:3}) we estimate an upper bound of the single photon coupling rate of approx. $2$\,mHz. This 
The small value of $g_0$ is consistent with our observation that the mechanical linewidth does not yield any measurable detuning dependence in our experiment (not shown).

In order to obtain a better estimation of the electromechanical coupling strength, we apply a second technique to characterize the parametric coupling between the nanostring and the three-dimensional cavity. It is based on the optomechanically induced transparency (OMIT),~\cite{Agarwal2010,Weis2010,Teufel2011} which arises from the coherent interaction of two microwave tones with the mechanical resonator in the resolved sideband regime.~\cite{Singh2014} In an OMIT experiment, the cavity is strongly pumped by a drive tone $\omega_d=\omega_c-\Omega_m$ which is red detuned from the cavity resonance by the frequency of the mechanical resonator. The cavity response is measured by a second, weak probe tone $\omega_p$ that is scanned across the cavity resonance. The beating between the two microwave tones induces a radiation pressure force coherently driving the mechanical resonator. In turn, the resonator imprints sidebands on the drive which interfere constructively with the probe. This opens a transparency window in the cavity transmission (or, as in our case, the cavity reflection which strictly speaking leads to optomechanically induced reflection (OMIR)~\cite{Singh2014}). According to the standard theory of OMIT, the height of the transparency peak allows to directly extract the cooperativity $C=4g^2/(\kappa \Gamma_m)$, rendering OMIT an important tool in cavity opto- and electromechanics. 

In Fig.~\ref{fig:Fig_4}(a) we plot the response of the microwave cavity as a function of the probe frequency $\omega_p$ for a drive applied at optimal detuning $\Delta = \omega_d - \omega_c = -\Omega_m$. The drive and probe power are $15$\,dBm and $-25$\,dBm, respectively. A transparency feature at $\omega_c$ is clearly apparent in the center of the cavity resonance. Note that the peak is not symmetric. Its asymmetry reflects the nonlinear response of the mechanical resonator to the radiation pressure drive exerted by the two microwave tones.~\cite{Zhou2013,Shevchuk2015,Singh2016} In agreement with the standard model for OMIR for the case of an undercoupled cavity driven on the red sideband, the shape of the nonlinear transparency peak directly follows the response of the Duffing resonator.~\cite{Zhou2013,Shevchuk2015}
This is also reflected in the hysteretic behavior of the OMIR feature displayed in Fig.~\ref{fig:Fig_4}(b) for a forward (black) as well as a reverse (red) sweep of the cavity probe at a somewhat lower probe power to maximize the magnitude of the OMIR feature.
%\addAT{[The probe power in this case is lower than in Fig. 4a (exact number yet unknown. Lower probe power gives higher peak than in the high probe power case]}. 
%Clearly, a hysteretic behavior is found which reflects the Duffing response of the strongly driven mechanical resonator. 
Figure~\ref{fig:Fig_4}(c) illustrates the dependence of OMIR peak on the probe power. The curves (from left to right) correspond to an increasing probe power from $-46$\,dBm to $-39$\,dBm. As expected, the width of the nonlinear feature broadens with increasing probe power while its amplitude slightly decreases.\cite{Shevchuk2015} Finally, the effect of a detuning $\delta$ of the cavity drive from the red sideband condition, $\Delta = -\Omega_m + \delta$, is explored in Fig.~\ref{fig:Fig_4}(d) for a constant probe 
%and drive 
power of $-25$\,dBm. %and $15$\,dBm, respectively. 
For a drive tone red- or blue-shifted from the red-sideband condition ($\delta = \pm 180$\,kHz, red and blue trace, respectively), the OMIR peak moves away from the minimum of the cavity response at $\omega_c$.~\cite{Shevchuk2015} The data for the drive matched to the red sideband is also included ($\delta=0$\,Hz, black trace). For clarity, the red and blue trace are vertically offset from the black trace.

According to the standard theory of OMIR, the height of the transparency peak $S_{11}^0$ allows to quantify the cooperativity. For the case of an undercoupled cavity driven on the red sideband, $C = (2\eta)/(1-S_{11}^0)-1$. Neither the nonlinear regime nor a poor sideband resolution greatly affect the magnitude of the OMIR feature.~\cite{Shevchuk2015,Bodiya2019} Using the yellow trace in Fig.~\ref{fig:Fig_4}(c), extract an approximate value of $C\approx 0.025$. This translates into an optomechanical coupling strength of $g/(2\pi)=1.2$\,kHz, and, given a cavity photon number of $4.0\cdot 10^{13}$ at a drive power of $15$\,dBm on the red sideband, to a single photon coupling rate of $g_0/(2\pi) = 200\,\mu$Hz.

It is noteworthy to mention that such a feeble single photon coupling can produce observable features in the OMIT experiment at all. This is enabled by the large number of photons supported by the three-dimensional microwave cavity, which exceeds photon numbers achieved in planar microwave resonators at low temperatures~\cite{Singh2014,Teufel2011} by at least four orders of magnitude and is at present only limited by the maximum output power of our microwave generator.
%[VERGLEICH MIT BENCHMARKS; in meinem Vortrag hab ich mir als Größenordnungen für n in optischen cavities 1e3 und in muw cavities 1e8 notiert, aber ohne Referenzen.]
The observed weak single photon coupling strength is attributed to our cavity design, as the loop antenna could not be precisely positioned in the cavity in our experiment and presumably only weakly couples to the $\textrm{TM}_{110}$ cavity mode. At the same time, the electromechanical coupling of the nanostring to the electrodes is limited by a relatively large electrode-electrode separation of approx. $600$\,nm for the sample under investigation. For future work on the room temperature cavity electromechanics platform, an improved control of the antenna position as well as a smaller electrode gap will enable to significantly increase $g_0$.

%\addAT{We attribute the strongly suppressed OMIR peak and the corresponding underestimation of the cooperativity to a large population of approx. $7500$ thermal photons in the cavity, which is not accounted for in the standard theory. We further hypothesize that the reduced magnitude of the OMIR peak can be employed to independently probe the number of thermal photons in the cavity, which may also be of interest for cavity electromechanics at cryogenic temperatures where a small but finite thermal cavity population typically exists.}
\section{Conclusion}
In conclusion, we have demonstrated a cavity electromechanical system operating in the sideband resolved regime at room temperature. This was accomplished by introducing a three-dimensional, non-superconducting microwave cavity made of copper which replaces the previously employed copper microstrip cavity, the quality factor of which is outperformed by more than an order of magnitude. In our experiment a non-metallized silicon nitride nanostring resonator was dielectrically coupled to the $\textrm{TM}_{110}$ mode of the cavity which offers almost perfect sideband resolution $\kappa \leq \Omega_m$. Electromechanical coupling was observed and characterized in one- and two-microwave tone experiments. While the mechanical eigenfrequency shift is dominated by dielectric frequency tuning, the optomechanically induced transparency (in reflection geometry) establishes a clear proof of dynamical backaction. Despite the minute single photon coupling rate of our first implementation of the room temperature cavity electromechanical plattform in the sub-mHz regime, a measureable coupling is enabled by the large number of photons circulating in the three-dimensional microwave cavity. As a result of the required strong red-detuned cavity drive, the response of the mechanical resonator is nonlinear in our proof-of-principle experiment. 

Our results translate the thriving field of cavity electromechanics from the millikelvin realm to room temperature. For future exploitation, the electromechanical vacuum coupling rate $g_0/(2\pi)$ will need to be increased. This can be accomplished by an improved positioning of the loop antenna providing the coupling between the cavity mode and the control electrodes. Furthermore, the coupling can be enhanced by increasing the dielectric transduction efficiency, i.e. by reducing the lateral gap between the electrodes. Following these technical improvements, we expect to reveal the  electromechanical cooling or pumping of the mechanical mode. Finally, the quality factor or the $\textrm{TE}_{011}$ mode exceeds that of the $\textrm{TM}_{110}$ mode by another order of magnitude, which offers the prospect of entering the deep-sideband-resolved regime of cavity electromechanics at room temperature.

\begin{acknowledgments}
Financial support from the Deutsche Forschungsgemeinschaft 
(DFG, German Research Foundation) via Project-ID 425217212 (Collaborative Research Center SFB 1432)); the European Unions Horizon 2020 programme for Research
and Innovation under grant agreement No. 732894 (FET Proactive HOT), as well as the German Federal Ministry of Education and Research through contract no. 13N14777 funded within the European QuantERA cofund project QuaSeRT is gratefully acknowledged. 
\end{acknowledgments}
\section*{Data Availability}
The data that support the findings of this study are available from the corresponding author upon reasonable request.

\section*{References}
%\nocite{*}
\bibliography{3D_Bib}% Produces the bibliography via BibTeX.

\end{document}